\documentstyle[aps,12pt,epsfig]{revtex}
\newcommand{\gmu}{{\gamma_\mu}}
\newcommand{\gf}{{\gamma_5}}
\newcommand{\tauiso}{{\mbox{\boldmath $\tau$}}}
\newcommand{\bm}{\bibitem}
\begin{document}
\draft
\tighten
\title{Role of baryonic resonances in the dilepton emission in
nucleon-nucleon collisions}
\date{\today}
\author{R. Shyam\footnote{On leave from Saha Institute of Nuclear Physics,
Calcutta, India}
 and U. Mosel}
\address { 
Institut f\"ur Theoretische Physik, Universit\"at Giessen, D-35392 Giessen, 
Germany } 
\maketitle

\begin{abstract}
Within an effective Lagrangian model, we present calculations  
for cross sections of the dilepton production
in proton-proton and proton-neutron collisions at laboratory 
kinetic energies in 1-5 GeV range. Production amplitudes
include contributions from the nucleon-nucleon bremsstrahlung
as well as from the mechanism of excitation, propagation, and radiative
decay of $\Delta$(1232) and N$^*$(1520) intermediate baryonic
resonances. It is found that the delta isobar terms dominate
the cross sections in the entire considered beam energy range.
Our calculations are able to explain the data of the DLS 
collaboration on the dilepton production in proton-proton
collisions for beam energies below 1.3 GeV. However, for
incident energies higher than this the inclusion of contributions
from other dilepton sources like Dalitz decay of $\pi^0$ and
$\eta$ mesons, and direct decay of $\rho^0$ and $\omega$ mesons
is necessary to describe the data. 
\end{abstract}
 
\newpage
\section{Introduction}
Dileptons observed in the nucleus-nucleus collisions 
travel relatively unscathed from the production point to the detector.
Therefore, they are expected to provide clear information about the
early dense and hot stage of heavy ion collisions~\cite{gal87,xia88,wol90},
which is in contrast to the hadronic probes which often suffer from strong
final state interactions and the information about the collision history
carried by them may be lost due to the rescattering in the
expansion phase~\cite{mos91}.  One of the phenomena 
predicted~\cite{bro91,hat92,cha92,her93,asa93,sak94,kli96,rap97,leu98}
at higher nuclear matter densities is the restoration of chiral symmetry
which is manifested in the modification of masses of the vector mesons as a
function of the nuclear matter density. Consequences of this effect can be
observed in the dilepton ($e^+e^-$) spectra measured in nucleus-nucleus
collisions~\cite{wol93}. Enhancements (above known sources) observed in   
the measured~\cite{aga95,ake95} soft lepton pair production cross sections
in heavy ion collisions at the SPS energies, have been 
attributed~\cite{lkb95,klb96,cas95,bra97,cas98} to in-medium modifications
of the vector meson properties rather than to reflection of a new state
of hadronic matter.  Although, such a scenario has not been fully
successful~\cite{bra98,ern98,brk99} in explaining the dilepton spectra
measured in these reactions at much lower beam energies
(a few GeV/nucleon)~\cite{por97} where the temperature and density
regime is different and one does not expect a phase transition.

In this context, the 
investigation of the dilepton production in elementary nucleon-nucleon
($NN$) collisions is of interest because the corresponding cross sections 
enter in the transport model calculations of the $e^+e^-$
spectra in the heavy ion collisions. Therefore, a quantitative understanding
of this process is a natural prerequisite to an unequivocal determination 
of in-medium effects mentioned above. The study of this process is of 
interest in its own right as it is expected to provide deeper insight
into the hadronic structure and the photon-hadron interactions. This 
investigation may give fundamental information on the electromagnetic form
factor of the nucleon in the time-like region around the vector-meson
masses which is otherwise hard to access(see, e.g.,~\cite{mos91}).

It has been known for some time that intermediate baryonic resonances play
an important role in the dilepton production in the $NN$ collisions. 
Dalitz decay of the $\Delta$ isobar has been shown to be a strong
dilepton production channel~\cite{gal89,hag94,sch94}. The importance
of the baryonic resonance N$^*$(1520) has been emphasized for this
process in Refs.~\cite{bra99,bra01} where it is pointed out that
the subthreshold $\rho$ production (and its subsequent decay) via
this resonance makes important contributions to the dilepton spectrum
observed in the proton-proton ($pp$) collisions. Therefore, the
investigation of the dilepton production is  expected to provide
a useful tool to probe the parameters ({$\it e.g.$}, coupling constants,
form factors etc.) of the nucleon-resonance-photon vertices and masses
and widths of the relevant nucleon resonances.

The aim of this paper is to investigate the dilepton production in  
nucleon-nucleon collisions in the beam energy range of 1-5 GeV
within an effective Lagrangian model (ELM) which has been successfully
applied to the description of pion and associated kaon production in
$pp$ collisions~\cite{eng96,shy98,shy99}. 
Initial interaction between two incoming nucleons is modeled by an
effective Lagrangian which is based on the exchange of the $\pi$,
$\rho$, $\omega$ and $\sigma$ mesons. The coupling constants at the
nucleon-nucleon-meson vertices are determined by directly fitting
the $T$ matrices of the nucleon-nucleon ($NN$) scattering
in the relevant energy region. The effective Lagrangian  
uses the pseudovector (PV) coupling for the nucleon-nucleon-pion
vertex and thus incorporates the low energy theorems of current
algebra and the hypothesis of PCAC. The $e^+e^-$ production proceeds
via excitation, propagation and radiative decay of $\Delta$(1232) and
$N^*$(1520) baryonic resonance states. Also included are the nucleon
intermediate states (which gives rise to the $NN$ bremsstrahlung contribution).
The interference terms between various amplitudes are taken into account. 
The gauge invariance at the electromagnetic vertices 
is preserved in our calculations. Our model is similar in spirit
to those of Refs.~\cite{sch94,fdj96,hag91}. However,
in Ref.~\cite{hag91} no resonance contribution was considered whereas
in Ref.~\cite{sch94,fdj96} resonance 
contributions were limited to the $\Delta$ isobar only. The
latter, though, did include the interference between the nucleon
and $\Delta$ terms. In calculations presented in
Refs.~\cite{bra99,bra01,bra95}, $NN$ bremsstrahlung contributions
were ignored. We would like to stress that in Refs.~\cite{bra99,bra01}
constant matrix elements have been used for various processes
and the total dilepton production cross sections have been calculated
by adding the corresponding cross sections and not the amplitudes, so 
part of the motivation of the present study is the investigation
of the so far neglected quantum mechanical effects.

We investigate the role of baryonic resonances in the  
invariant mass spectrum of the dilepton produced in proton-proton and
proton-neutron collisions at various beam energies in the 
1-5 GeV range.  To this end, we present the first field theoretic
calculation of the dilepton production in $NN$ collisions where
excitation, propagation and radiative decay of the N$^*$(1520)
baryonic resonance are fully accounted for. We also compare our
calculations with the published data~\cite{wil98} on the dilepton
production in elementary proton-proton collisions, by the Dilepton
Spectrometer (DLS) collaboration. In order to describe these data the
contributions from other dilepton sources
($\pi^0$ and $\eta$ Dalitz decay and direct decay of $\rho^0$ and
$\omega$ mesons) have also been considered. 
 
The remainder of this paper is organized in the following way. 
Section II contains details of our theoretical approach. Section
III comprises the results of our analysis and their discussions. 
The summary and conclusions of our work is presented in Sec. IV.
 
\section{FORMALISM}  
 
A representative of the lowest order Feynman diagrams contributing to
the dilepton production as considered by us, is shown in Fig.~1.
The intermediate nucleon or resonances can radiate a virtual photon
which decays into a dilepton (Figs.~1a and 1b). There are also their
exchange counterparts. In addition, there are diagrams of these
types where the virtual photon is emitted from the nucleon line on the
right side.  The internal meson lines can also lead to
dilepton emission (see Fig 1c). Momenta of various particles are
indicated in Fig.~1a. $q$, $p_i$, and $k$ are four momenta of the 
exchanged meson, the intermediate resonance (or nucleon) and the photon,
respectively. To evaluate various amplitudes, 
we have used the effective Lagrangians for the nucleon-nucleon-meson,
resonance-nucleon-meson, nucleon-nucleon-photon and
resonance-nucleon-photon vertices. These are discussed in the 
following subsections.
\begin{figure}[here]
\begin{center}
\mbox{\epsfig{file=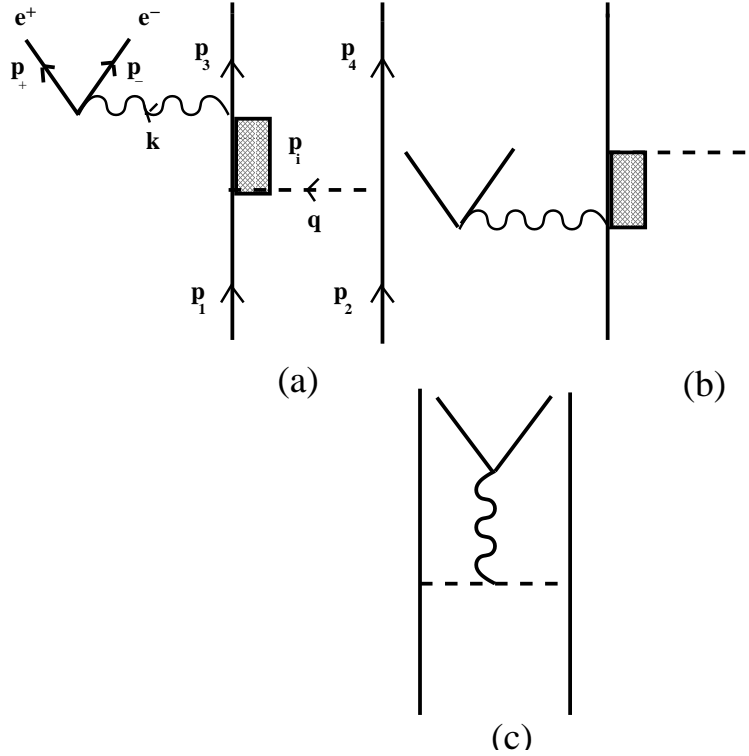,height=10cm}}
\end{center}
\vskip .1in
\caption{
A representative of Feynman diagrams for emission of dilepton
in nucleon-nucleon collision as considered in this work.
(a) denotes emission after $NN$ collisions, (b) before $NN$ collision
and (c) during $NN$ collision. The box represents any of an off-shell
nucleon, a $\Delta$ isobar or a $N^*$ resonance. 
}
\label{fig:figa}
\end{figure}
 
\subsection{Nucleon-nucleon-meson vertex}

As done before in the investigation of $pp \rightarrow pp\pi^0$,
$pp \rightarrow pn\pi^+$~\cite{eng96} and
$pp \rightarrow p \Lambda K^+$~\cite{shy99} reactions, the parameters
for these vertices are determined by fitting the $NN$ elastic scattering
$T$ matrix with an effective $NN$ interaction based on the
$\pi$, $\rho$, $\omega$, and $\sigma$ meson exchanges. The effective
meson-$NN$ Lagrangians are (see, e.g.,~\cite{wis88,pent02})
 
\begin{eqnarray}
{\cal L}_{NN\pi} & = & -\frac{g_{NN\pi}}{2m_N} {\bar{\Psi}}_N \gamma _5
                             {\gamma}_{\mu} \tauiso
                            \cdot (\partial ^\mu {\bf \Phi}_\pi) \Psi _N. \\
{\cal L}_{NN\rho} &=&- g_{NN\rho} \bar{\Psi}_N \left( \gmu + \frac{k_\rho}
                         {2 m_N} \sigma_{\mu\nu} \partial^\nu\right)
                          \tauiso \cdot \mbox{\boldmath $\rho$}^\mu \Psi_N. \\
{\cal L}_{NN\omega} &=&- g_{NN\omega} \bar{\Psi}_N \left( \gmu + \frac{k_\omega}
                         {2 m_N} \sigma_{\mu\nu} \partial^\nu\right)
                          \omega^\mu \Psi_N.   \\
{\cal L}_{NN\sigma} &=& g_{NN\sigma} \bar{\Psi}_N \sigma \Psi_N.
\end{eqnarray} 
In Eqs.~(1)-(4), we have used the notations and conventions of
Bjorken and Drell~\cite{bjo64} and definitions of various terms
are the same as those given there. In Eq.~(1) $m_N$ denotes
the nucleon mass.
It should be noted that we use a PV coupling for the $NN\pi$ vertex.
Since these Lagrangians are used to directly model the $NN$ $T$ matrix, we 
have also included a nucleon-nucleon-axial-vector-isovector vertex,
with the effective Lagrangian given by
\begin{eqnarray}
{\cal L}_{NNA} & = & g_{NNA} {\bar {\Psi}} \gamma_5 \gamma_\mu \tauiso \Psi
                     \cdot {\bf {A}}^\mu,
\end{eqnarray}
where ${\bf A}$ represents the axial-vector meson field.
This term
is introduced because if the mass of the axial meson $A$ is chosen to
be very large ($\gg m_N$)~\cite{sch94} and $g_{NNA}$ is defined as
\begin{eqnarray}
g_{NNA} & = & \frac{1}{\sqrt{3}} m_A \left(\frac{g_{NN\pi}}{2m_N}\right),
\end{eqnarray}
it cures the unphysical behavior in the angular distribution of
$NN$ scattering caused by the contact term in the one-pion exchange
amplitude. It should be mentioned here that $A$ is different from
the $a_1$(1260) meson resonance to be discussed later on. The role of the
$A$ vertex is to explicitly subtract out the contact term of the one-pion
exchange part of the $NN$ interaction. Similar term in the
coordinate space potential is effectively switched off by the repulsive
hard core.
 
At each interaction vertex, the following form factor is introduced 
\begin{eqnarray}
F_{i}^{NN} & = & \left (\frac{\lambda_i^{2} - m_i^{2}}{\lambda_i^{2} - q_i^{2}}
        \right ), i= \pi, \rho, \sigma, \omega,
\end{eqnarray}
where $q_i$ and $m_i$ are the four momentum and mass of the $i$th
exchanged meson and $\lambda_i$ is the corresponding cut-off parameter.
The latter governs the range of suppression of the contributions of
high momenta which is done via the form factor. Since $NN$ scattering
cross sections decrease gradually with the beam energy (beyond certain value),
and since we fit the elastic $T$ matrix directly, the coupling
constants are expected to be energy dependent. Therefore, while keeping the
cut-offs $\lambda_i$ [in Eq.~(7)] energy independent, 
we take energy dependent meson-nucleon coupling constants of the following
form 
\begin{eqnarray}
g(\sqrt{s}) & = & g_{0} exp(-\ell \sqrt{s}),
\end{eqnarray}
here $s$ is the square of the total CM energy. The
parameters $g_0$, $\ell$, and $\lambda$ were determined by fitting to
the $T$ matrix to the relevant
proton-proton and proton-neutron scattering data at the 
beam energies in the range of 800 MeV to 4.00 GeV~\cite{sch94}. This 
procedure also fixes the sign of the effective Lagrangians [Eqs. (1)-(5)].
The values of various parameters are shown in Table I [the signs of all the
coupling constants (g) are positive]. In this table the parameters of the
$A$ exchange vertex are not explicitly shown as they are related to those
of the pion via Eq.~(6). The mass of the $A$ meson is taken to be 18 GeV.
The same parameters were also
used to describe the initial $NN$ interaction in
the calculations reported in Refs.~\cite{sch94,eng96,shy99}.
This ensures that the elastic $NN$ elastic scattering channel
remains the
\begin{table}[here]
\begin{center}
\caption[T1]{ Coupling constants for the $NN$-meson vertices used in the
calculations}
\vspace{.5cm}
\begin{tabular}{|c|c|c|c|c|}
\hline
 Meson & $g^2/4\pi$ & $\ell$ & $\Lambda$ & mass \\
       &             &        & (\footnotesize{GeV} ) & (\footnotesize{GeV})
 \\ \hline
$\pi    $ & 12.562 & 0.1133 & 1.005 & 0.138 \\
$\sigma $ & 2.340  & 0.1070 & 1.952 & 0.550 \\
$\omega $ & 46.035 & 0.0985 & 0.984 & 0.783 \\
$\rho   $ & 0.317  & 0.1800 & 1.607 & 0.770 \\
$k_{\rho}$ = 6.033, $k_{\omega}$ = 0.0 & & & & \\ \hline
\end{tabular}
\end{center}
\end{table}
\noindent
same in the description of various inelastic processes 
within this model, as it should be.

The exchange of $a_1$ meson (mass = 1260 MeV), which is the chiral partner
of the $\rho$ meson has also been considered by some authors~\cite{dur84}
in the description of the $NN$ interaction.
However, its  contribution
there is masked to a considerable extent by the strong short-ranged repulsion
originating from $\omega$ exchange. Moreover, as can be seen from Table I,
the values of the cut-off parameters appearing in the form factors
[Eq.~(7)] are in the region of 1.0 - 2.0 GeV. Therefore, it does not seem  
meaningful to include exchange of mesons with masses lying in
the region where modifications due to the extended structure of hadrons 
are applied. Nevertheless, $a_1$ meson (if included) can contribute 
to the dilepton spectrum through diagrams like the one shown in Fig.~1(c).
However, the couplings of this meson to nucleons and resonances are
largely unknown. In any case, contributions of such processes
are expected
to be of relevance only for the dilepton invariant masses
$>$ 1.5 GeV~\cite{chu94}. This region is excluded by the DLS data.
Due to these reasons we have not included the exchange of $a_1$ meson
in our calculations. 

\subsection{Resonance-nucleon-meson vertex}

In addition to nucleonic intermediate states ($NN$ bremsstrahlung), we
have considered in this work also the  contributions from the 
$\Delta$(1232) isobar and $N^*$(1520) nucleon resonance intermediate states.  
The latter is a spin-${3 \over 2}$ negative parity resonance.
While only $\pi$ and $\rho$ mesons couple to $\Delta$ isobar, all
of the four exchanged
mesons, namely, $\pi$, $\rho$, $\omega$, and $\sigma$ can couple 
to $N^*$(1520) resonance. 

For spin-${3 \over 2}$
 resonances, we use the following effective
Lagrangians~\cite{feu97}
\begin{eqnarray}
{\cal L}_{RN\pi} & = & \frac{g_{RN\pi}}{m_\pi} {\bar{\Psi}}^R_\mu
                         \Gamma_\pi \partial^{\mu}
                         {\bf \Phi}_\pi \Psi _N + {\rm H.c.}. \\
{\cal L}_{RN\rho} & = & {\rm i} \frac{g_{RN\rho}}{m_\rho}
                        {\bar{\Psi}}^R_\mu\left(
                        \partial^\nu \mbox{\boldmath $\rho$}^\mu -
                        \partial^\mu \mbox{\boldmath $\rho$}^\nu
                        \right) \gamma_\nu \Gamma \Psi _N + {\rm H.c.}.\\
{\cal L}_{RN\omega} & = & {\rm i} \frac{g_{RN\omega}}{m_\omega}
                        {\bar{\Psi}}^R_\mu \left(
                        \partial^\nu \omega^\mu - \partial^\mu \omega^\nu
                        \right) \gamma_\nu \Gamma \Psi _N + {\rm H.c.}.\\
{\cal L}_{RN\sigma} & = & \frac{g_{RN\sigma}}{m_\sigma}
                         {\bar{\Psi}}^R_\mu \Gamma_\pi (\partial ^{\mu}
                         {\sigma}) \Psi _N + {\rm H.c.}. 
\end{eqnarray}
Here, ${\bar {\Psi}}^R_\mu$ is the vector spinor for the
spin-${3 \over 2}$
particle. In Eqs.~[(9) - (12)], the operator $\Gamma_\pi$
\begin{table}[here]
\begin{center}
\caption[T1]{Coupling constants and cut-off parameters for the
resonance-nucleon-meson vertices used in the calculations}
\vspace{.5cm}
\begin{tabular}{|c|c|c|c|} \hline
Resonance     & decay channel  &  $g$  & cut-off parameter \\
              &                &       & (\footnotesize{GeV})\\ \hline
$\Delta$(1232)& $N\pi $        & 2.13  & 1.421 \\
              & $N\rho$        & 7.14  & 2.273 \\
$N^*$(1520)   & $N\pi $        & 1.55  & 0.800 \\
              & $N\rho$        & 6.44  & 0.800 \\
              & $N\omega$      & 3.42  & 0.800 \\
              & $N\sigma$      & 1.24  & 0.800 \\
\hline
\end{tabular}
\end{center}
\end{table}
\noindent
is unity for even
parity resonance and $\gf$ for the odd parity one, whereas
$\Gamma$ is $\gf$ for the even parity resonance and unity
for the odd parity one.
The meson fields in above equations need to be 
replaced by $\tauiso \cdot\phi$ (where $\phi$ corresponds
to $\pi$, $\rho$, $\omega$, or $\sigma$ meson fields) and
${\bf T}\cdot \phi$ (where ${\bf T}$ and $\tauiso$ are the isospin
transition operator and Pauli isospin matrices, respectively) for   
isospin-${1 \over 2}$ and isospin-${3 \over 2}$
resonances, respectively. 
The values of the coupling constants $g_{\Delta N\pi}$ and
$g_{\Delta N\rho}$ are shown in Table II; these are
the same as those determined in Ref.~\cite{sch94} by fitting
to experimental data on $pp \rightarrow n\Delta^{++}$ reaction 
at kinetic energies in the range of 1-2 GeV. We have assumed that
the off-shell
dependence of the resonance-nucleon vertex is determined
solely by multiplying the vertex
constant by a form factor. Similar
to Ref.~\cite{sch94,eng96}, we have used the following  form factor
for the $\Delta$ vertices 
\begin{eqnarray}
F_{i}^{N\Delta} & = & \left [\frac{{(\lambda_i^\Delta)^2} - m_i^{2}}
{{(\lambda_i^\Delta)^2} - q_i^{2}} \right ]^2, i= \pi, \rho,
\end{eqnarray}
with the values of the cut-off parameters as $\lambda_\pi^\Delta$ 
and $\lambda_\rho^\Delta$ as given in Table II which are exactly the same
as those used earlier~\cite{sch94,eng96,shy99}.

For the $N^*N\pi$ and $N^*N\rho$ vertices the coupling constants have been 
determined from the observed branching ratios for the decay of the
resonance to $N\pi$ and $N\rho$ channels, respectively. In the later
case the finite lifetime for the decay $\rho \rightarrow \pi\pi$ has
been taken into account by introducing an integration over the
corresponding phase space. The details of this method are provided
in Ref~\cite{shy99}. The coupling constants $g_{N^*N\omega}$ and
$g_{N^*N\sigma}$ are determined by vector meson dominance (VMD)
hypothesis~\cite{pos01} and from the branching ratio of the decay of
this resonance into the two pion channel~\cite{shy99}, respectively.
It should be noted that there is considerable uncertainty in the latter
two coupling constants. However, the contributions of these terms to the
total dilepton production amplitude are almost negligible.
As branching ratios determine only the square of the corresponding
coupling constants, their signs remain uncertain in this method.
Predictions from the independent calculations can, however, 
be used to constrain these signs~\cite{feu97,man92,cap94}. Guided by
the results of these studies, we have chosen the positive sign for the
coupling constants for these vertices. The values of various coupling
constants are given in Table II.
   
As in Refs.~\cite{feu97,pos01}, we have used the following
form factor for the $N^*N$ vertices   
\begin{eqnarray}
F_{j}^{NN^*} & = & \left [\frac{{(\lambda_j^{N^*})^4}}
{{(\lambda_j^{N^*})^4} + (q_j^2-m_j^2)^2} \right ],
 j= \pi, \rho,\omega,\sigma,
\end{eqnarray}
with the value of the cut-off parameter being 0.8 GeV in all the cases.
It may however be mentioned that identical results will be obtained
if one uses the form factor as given by Eq.~(13) with different 
cut-off parameters.

\subsection{Nucleon-nucleon-photon vertex}

In the nucleonic bremsstrahlung process of dilepton production the 
intermediate nucleon is necessarily off-shell. The general form of
the effective Lagrangian is given by
\begin{eqnarray}
{\cal L}_{NN\gamma} & = & - e {\bar{\Psi}} \Gamma_\mu^{NN\gamma} \Psi A^\mu,
\end{eqnarray}
where the half-off-shell nucleon-photon vertex function $\Gamma_\mu$ 
is~\cite{kor95,nau87,tie90}
\begin{eqnarray}
\Gamma_\mu^{NN\gamma}& = -ie& \sum_{s=\pm} (F_1^s \gamma_\mu + F_2^s
 \Sigma_\mu + F_3^s k_\mu)
                  \Lambda_s.
\end{eqnarray}
In Eq.~(16), $k^2 = (p_f-p_i)^2$ with $p_f$ and $p_i$ being the
initial and final
nucleon four momenta. $\Lambda_\pm = (\pm p\!\!\!/+W)/W$ are the projection
operators where $W = (p^2)^{1/2}$ and  $\Sigma_\mu$ =
$i\sigma_{\mu \nu}k^\nu/2m_N$. The form
factors $F_{1,2,3}$ are functions of $k^2$, $W$, and $m_N$. The factor 
$F_3$ is not independent but is constrained by the Ward-Takahashi 
identity~\cite{itz64} (i.e. by the requirement of gauge invariance)
\begin{eqnarray}
F_1^\pm & = & {\hat{e}}_N + \frac{k^2}{\pm W -m_N}F_3^\pm,
\end{eqnarray}
where ${\hat{e}}_N$ is the nucleon charge in units of
$\mid e \mid$.

In this work, we have adopted the procedure followed in
Refs.~\cite{sch94,hag89} where one uses the on-shell form of the 
vertex function also for the off-shell momenta. This means, one assumes
$F_1^+ = F_1^- = F_1$ and $F_2^+ = F_2^- =F_2$ and $F_3^+ = F_3^- = 0$.
This vertex does not in general satisfy gauge invariance. However,
this can be achieved (see also~\cite{tow87}) by multiplying the external
photon emission vertices by the same form factors that multiply the
hadronic vertices [Eq.~(7)] and by multiplying the internal photon
production diagrams by the following additional factor:
\begin{eqnarray}
F_{int} & = & 1 + \frac{m_{meson}^2-q_1^2}{\lambda^2-q_2^2} + 
                  \frac{m_{meson}^2-q_2^2}{\lambda^2-q_1^2}, 
\end{eqnarray}
where $q_1$ and $q_2$ are the four momentum transfers at the 
left and right vertices, respectively. In this way various 
electromagnetic form factors can be implemented for the hadrons
without loosing gauge invariance.  Unfortunately, there are
problems and ambiguities in the the selection of form factors
$F_{1,2}$ (see $e.g$., ~\cite{pen02} for a detailed discussion).
It has been shown already in Ref.~\cite{sch94} that $NN$ bremsstrahlung
contributions depend sensitively on the choice of these form factors
and that cross sections calculated with no form factors are closest
to the data.  In our calculations, we  have used the
prescription~\cite{feu97,pen02} of using no form factors at the
electromagnetic vertices of the nucleon term and replacing
$F_1$ by the nucleon charge and $F_2$ by nucleon anomalous
magnetic moment. 
  
\subsection{Resonance-nucleon-photon vertex}
 
\begin{table}[here]
\begin{center}
\caption[T1]{Coupling constants for the
resonance-nucleon-photon vertices used in the calculations. For $N^*$(1520)
proton couplings are given in the first line and neutron couplings in the 
second line.}
\vspace{.5cm}
\begin{tabular}{|c|c|c|c|} \hline
Resonance     & $g_1$  & $g_2$ & $g_3$ \\
              &        &       &       \\ \hline
$\Delta$(1232)& 5.416  & 6.612 & 7.0 \\
$N^*$(1520)   & 3.449  & 5.074 & 1.0 \\
              & -0.307 & 1.862 & 3.0 \\
\hline
\end{tabular}
\end{center}
\end{table}
\noindent
For spin-${3 \over 2}$
resonance-nucleon-$\gamma$ vertices, the form of the
vertex functions is 
\begin{eqnarray}
\Gamma^{\mu \nu}_{RN\gamma} & = & -i\frac{e}{2m_N} 
              \left [g_1(k^2)\gamma_\lambda + \frac{g_2(k^2)}{2m_N}p^i_\lambda 
               + \frac{g_3(k^2)}{2m_N}k_\lambda \right ]
                (-k^\nu g^{\mu \lambda} + k^\mu g^{\nu \lambda})Z 
\end{eqnarray}
where $Z$ is $\gf$ for the even parity resonance and unity
for the odd parity one.
Values of the coupling constants $g_1$, $g_2$, and $g_3$
used in our calculations are shown in Table III. The first
two are taken from~\cite{feu97}, while the last one from~\cite{pen97}.
The vertex function given by Eq.~(19)
fulfills gauge invariance~\cite{noz90} by the way of its construction.  
 
It should be noted that the vector spinor vertices [Eqs.~(9)-(12)
and (19)] should in addition be contracted by an off-shell
projector $\Theta_{\alpha \nu}(z)\,=\, g_{\alpha \nu} -
{1 \over 2}(1 + 2z)\gamma_\alpha \gamma_\nu$
where $z$ is the off-shell parameter~\cite{feu97,ben95}.
This operator describes the off-shell admixture of spin-${1 \over 2}$
fields~\cite{ben89}. The choice
of the off-shell parameter $z$ is arbitrary and it is treated as a
free parameter to be determined by fitting the data. However, 
recently, the authors of Ref.~\cite{pas98} have proposed a different
$\pi N\Delta$ interaction which leads to amplitudes where  
spin-${1 \over 2}$ components of the Rarita-Schwinger propagator
(see in section E) drop out, thus making the off-shell parameters
redundant (see, $e.g.$,~\cite{pen02,pas98} for further details).
The full implication of this prescription on observables calculated
within the effective Lagrangian model will be investigated in future.
In the present study we work with the couplings given by
Eqs.~(9)-(12) and (19).

\subsection{Propagators}

In the calculation of the amplitudes, the propagators for various
mesons and nucleon resonances are required. For pion, $\rho$ meson and
axial-vector mesons, they are given by
\begin{eqnarray}
G_\pi(q) & = & {i \over (q^2 - m_\pi^2)}\\
G_\rho^{\mu\nu}(q) & = & -i\left({g^{\mu\nu}-q^\mu q^\nu/q^2}
                          \over {q^2 - m_\rho^2} \right)\\
G_A^{\mu\nu}(q) & = & -i\left(\frac{g^{\mu\nu}}{q^2-m_A^2}\right)
\end{eqnarray}
In Eq.~(22), the mass of the axial meson is taken to be very large,
as the corresponding amplitude is that of the contact term. The propagators
for $\omega$ and $\sigma$ mesons are similar to those given
by Eqs.~(21) and (20), respectively.

The propagators for the nucleon and the spin-${3 \over 2}$ resonance are
\begin{eqnarray}
G_{N}(p_i) & = & \frac{i(p_i\!\!\!\!\!/ + m_N)}{p_i^2 - m_{N}^2}
\end{eqnarray} 
 
\begin{eqnarray}
G_{R}^{\mu \nu} (p_i) & = & -\frac{i(p_i\!\!\!\!\!/ + m_R)}
                    {p_i^2 - [m_{R}-i(\Gamma_{R}/2)]^2} \nonumber \\
                 & \times & \left[ g^{\mu \nu} - \frac{1}{3}
     \gamma^\mu \gamma^\nu - \frac{2}{3m_{R}^2} p_i^\mu p_i^\nu
     + \frac{1}{3m_{R}^2} ( p_i^\mu \gamma^\nu - p_i^\nu \gamma^\mu ) \right].
\end{eqnarray}
In Eqs.~(24), $\Gamma_{R}$ is the total width of the resonance which
is introduced in the denominator term $(p^2-m_{R}^2)$ to account for the
fact that the resonances are not the stable particles; they
have a finite life time for the decay into various channels. $\Gamma_{R}$ 
is a function of the center of mass momentum of the decay channel, and it is
taken to be the sum of the corresponding widths for its decays to pion
and rho channels (the other decay channels are considered only
implicitly by adding their branching ratios to that of the pion channel):
\begin{eqnarray}
\Gamma_{R} & = & \Gamma_{R \rightarrow N\pi} +
                      \Gamma_{R \rightarrow N\rho}
\end{eqnarray}
The partial decay width $\Gamma_{R \rightarrow N\rho}$ is calculated
in the same way as in Ref.~\cite{shy99}. This method is also
identical to that used in Ref.~\cite{bra99}; the only difference is that
while Ref.~\cite{shy99} uses a fully relativistic expression, that 
employed in~\cite{bra99} is its non-relativistic counterpart. The
partial width $\Gamma_{R \rightarrow N\pi}$ is taken as~\cite{bra99}
\begin{eqnarray}
\Gamma_{R \rightarrow N\pi}(\mu) & = & \Gamma_0 \left( \frac{k_\pi(\mu)}
{k_\pi(m_R)} \right)^{2\ell+1} 
\left( \frac{0.25+k_\pi(m_R)^2}{0.25+k_\pi(\mu)^2} \right)^{2\ell+1}, 
\end{eqnarray}
with
\begin{eqnarray}
k_\pi(\mu) & = & \frac{[(\mu^2 - (m_\pi + m_N)^2)
(\mu^2 - (m_\pi - m_N)^2)]^{1/2}} {2\mu}. 
\end{eqnarray}
In Eq.~(26), $\ell$ and $\Gamma_0$ are 2 and 0.095 GeV for the  
$N^*$(1520) resonance, and 1 and 0.120 GeV for $\Delta$ isobar, respectively. 

It should be noted that no width is included in the $\rho$ propagator
[Eq.~(21)] as it corresponds to a exchange process between two nucleons
where $q^2$ is in the space-like region and hence $\rho$ meson has a
zero width. However, this width is finite in the $\rho$ propagator in,
e.g., $NN \rightarrow NN\rho \rightarrow NN e^+e^-$ process where the
corresponding momentum is time-like.
                
\subsection{Amplitudes and cross sections}

After having established the effective Lagrangians, coupling constants and
form of the propagators, we can now proceed to calculate the amplitudes
for various diagrams associated with the $NN \rightarrow NN e^+e^-$
reaction. These amplitudes can be written by following the well known 
Feynman rules~\cite{itz64} and calculated numerically. 
It should be stressed here that the signs of the various amplitudes
are fixed, by those of the effective Lagrangians, coupling constants
and the propagators as described above. These signs are not 
allowed to change anywhere in the calculations. 

The general formula for the invariant cross section of the $N + N
\rightarrow N + N + e^+e^-$ reaction is written as 
\begin{eqnarray}
d\sigma & = & \frac{m_N^4m_e^2}{2\sqrt{[(p_1 \cdot p_2)^2-m_N^4]}}
                     \frac{1}{(2\pi)^8}\delta^4(P_f-P_i)|A^{fi}|^2
                     \prod_{a=1}^4 \frac{d^3p_a}{E_a},
\end{eqnarray}
where $A^{fi}$ represents the total amplitude, $P_i$ and $P_f$
the sum of all the momenta in the initial and final states, respectively, and
$p_a$ the momenta of the particles in the final state. In Eq.~(28) $m_e$
represents the mass of the electron. The term $\mid A^{fi} \mid$
already includes a sum over final spin and average over initial spin
degrees of freedom of all particles. More details of the evaluation of
Eq.~(28) are given in appendix A.
\begin{figure}[here]
\begin{center}
\mbox{\epsfig{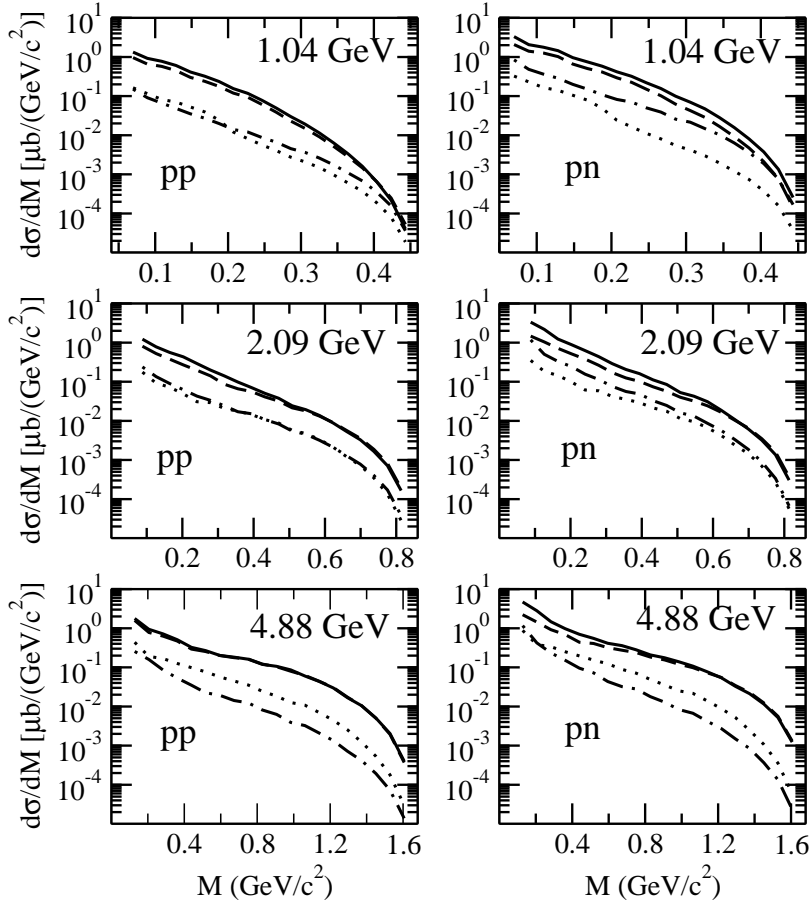}}
\end{center}
\vskip .1in
\caption {
Calculated invariant mass distributions for dileptons
produced in proton-proton (left panel) and proton-neutron 
collisions (right panel) at the beam energies of 1.04 GeV, 2.09 GeV and
4.88 GeV. The contributions of the $NN$ bremsstrahlung (non-resonance),
and the delta isobar and $N^*$(1520) resonance states are shown
by dashed-dotted, dashed, and dotted lines, respectively. Their
coherent sum is shown by solid lines. 
}
\label{fig:figb}
\end{figure}
\noindent
\section{Results and discussions}

In Fig.~2, we show the invariant mass spectra for dilepton production
in both $pp$ and $pn$ collisions at bombarding energies of 1.04 GeV,
2.09 GeV and 4.88 GeV. It can be noted 
that in all the cases the dominant contribution arrises from the
intermediate states consisting of
the $\Delta$ isobar resonance. In fact the total yields are almost equal to
the contributions of
the $\Delta$ amplitude alone. The $pn$ cross sections 
are about a factor of 2-3 larger than those for the $pp$ reactions
even at the higher beam energy of 4.88 GeV.
For beam energies of 1.04 GeV and 2.09 GeV, the contributions of the 
$N^*$(1520) terms are similar to those of the $NN$ bremsstrahlung
processes for the $pp$ collisions, while for the $pn$ case the latter
are larger than the former. However, at 4.88 GeV, the $N^*$ 
channel prevails over the bremsstrahlung one in both cases.
Any way, the $\Delta$ isobar production terms are still
dominant even at this high energy.
\begin{figure}[here]
\begin{center}
\mbox{\epsfig{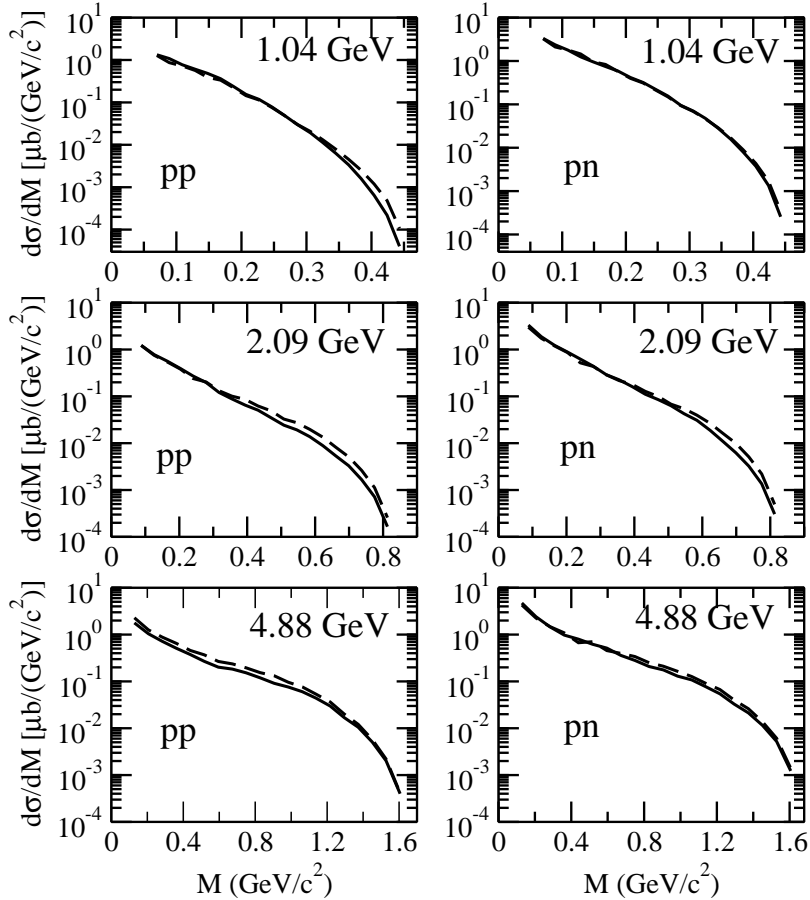}}
\end{center}
\vskip .1in
\caption {
Calculated invariant mass distributions for dileptons
produced in proton-proton (left panel) and proton-neutron 
collisions (right panel) at the beam energies of 1.04 GeV, 2.09 GeV and
4.88 GeV. Solid and dashed lines show the cross sections obtained by
coherent and incoherent summations of the amplitudes of various processes
shown in Fig.~2, respectively.
}
\label{fig:figc}
\end{figure}
\noindent
The contributions of the nucleon bremsstrahlung process (graphs
involving only intermediate nucleon lines) are considerably larger
in case of $pn$ reaction as compared to those for the $pp$ case  
for all the beam energies.  We do not observe the reversal
of this trend at 4.88 GeV as seen in the soft-photon model
calculations of Ref~\cite{hag94} (where $NN$ bremsstrahlung is 
larger for $pp$ reaction as compared to that for $pn$ one). On the
other hand, if the off-shell behavior of the effective $NN$
interaction is described by using a $T$ matrix which includes the
$\Delta$ degrees of freedom, instead of the meson exchange mechanism
as employed in our model, the $pp$ bremsstrahlung contributions 
turn out to be larger than those of ours at the
bombarding energy of 1 GeV.
Yet due to lack of calculations for the $pn$ case,
it may not be possible to make comments about the relative
$pp$ and $pn$ bremsstrahlung contributions within this approach. 
In any case, the coherent 
sum of the $\Delta$ and bremsstrahlung terms as obtained within our
model are similar to those calculated with the $T$ matrix method. 

The role of the interference effects between various terms is
investigated explicitly in Fig.~3 where we compare the
cross sections obtained by coherent and incoherent summations of the
amplitudes corresponding to various processes shown in Fig.~2. The
former are the same as those shown by solid lines in Fig.~2.
It is already noted in Fig.~2 that the delta
\begin{figure}[here]
\begin{center}
\mbox{\epsfig{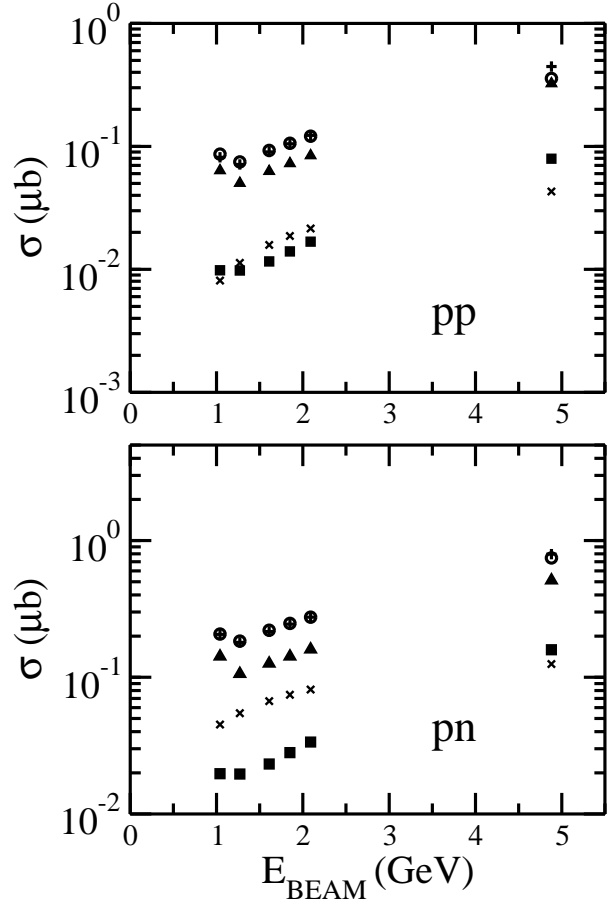}}
\end{center}
\vskip .1in
\caption {
Total dilepton production cross section in proton-proton
(top) and proton-neutron (bottom) collisions as a function of beam
energy. The nucleon, delta isobar and $N^*$(1520) contributions are
shown by crosses, filled triangles and solid squares, respectively. The
total cross sections obtained by coherent and incoherent summations of
the corresponding amplitudes are shown by open circles and plus signs,
respectively.
}
\label{fig:figd}
\end{figure}
\noindent
contributions are
slightly larger than the total ones at the larger mass end of the
spectrum for all the beam energies in both the cases. 
In Fig.~3, one can further see that the interference effects are noticeable
towards the larger mass ends of the spectra for beam energies below
4.88 GeV in case of $pp$ collisions.  For the 4.88 GeV case 
these effects show up also at lower values of the invariant mass.
On the other hand, they are relatively smaller for the $pn$ case
everywhere.
  
Contributions of various terms to the excitation function (integrated 
dilepton cross section) are shown in Fig.~4 for the $pp$ (top) and
$pn$ (bottom) collisions. The total cross section
at 1.04 GeV is
slightly  larger than that at 1.27 GeV  in both the cases. However,
after this it rises monotonously with the beam energy. The 
delta isobar terms dominate the total production yields for
all the beam energies. For $pp$ collisions, $NN$ bremsstrahlung and
$N^*$(1520) terms are of similar magnitude except at the beam energy of 
4.88 GeV where the latter term is larger than the former. However, 
in the $pn$ case, the $NN$ bremsstrahlung contributions are larger than
those of the $N^*$(1520) at lower beam energies while the two terms
contribute similarly at 4.88 GeV. Furthermore, the difference in the
interference effects of
\begin{figure}[here]
\begin{center}
\mbox{\epsfig{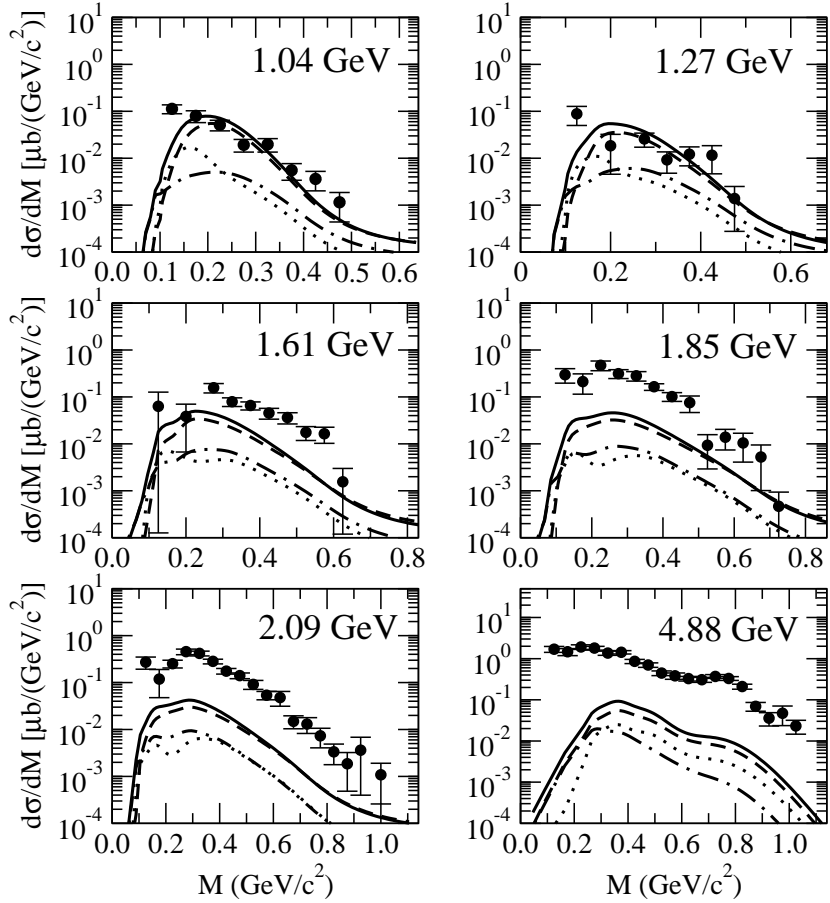}}
\end{center}
\vskip .1in
\caption {
Comparison of the effective Lagrangian model results for
the invariant mass spectra of dileptons produced in proton-proton
collisions, folded with the experimental filter,
with the data of the DLS collaboration~\protect\cite{wil98} at
the beam energies of
1.04 GeV, 1.27 GeV, 1.61 GeV, 1.85 GeV, 2.09 GeV and 4.88 GeV. 
Various curves have the same meaning as in Fig.~2. The solid
circles show the experimental data.
}
\label{fig:fige}
\end{figure}
\noindent
various terms in two case, should also be noted.
Cross sections obtained by coherent (shown by open circles) and incoherent
(shown by plus signs) summations of the amplitudes differ very slightly
for all the beam energies in case of the $pn$ collisions and at energies
smaller than 4.88 GeV for the $pp$ case. This can be understood from
the fact that the regions where the interference effects are visible in
Fig.~3, contribute very little to the integrated cross sections for 
these cases. However, the interference effects do show up even in the
total dilepton yields for $pp$ collisions at 4.88 GeV beam energy. 
 
The cross sections shown in Figs.~2 and 3 can not be compared 
data of the DLS collaboration. They have 
to be folded with appropriate experimental filter, which
is a function of the invariant mass ($M$), transverse momentum
($p_T$) and the rapidity in the laboratory frame ($y_{lab}$) 
of the produced dileptons. We have used the DLS acceptance filter
(version 4.1) in the folding procedure. In Fig.~5, we 
present a comparison of the folded invariant mass spectra (which also
include the final mass resolution~\cite{bra99}) and the data from
the DLS collaboration. We note that at the beam energies of 1.04 GeV and
1.27 GeV the effective Lagrangian model calculations are able to
describe the data reasonably well in the region of
\begin{figure}[here]
\begin{center}
\mbox{\epsfig{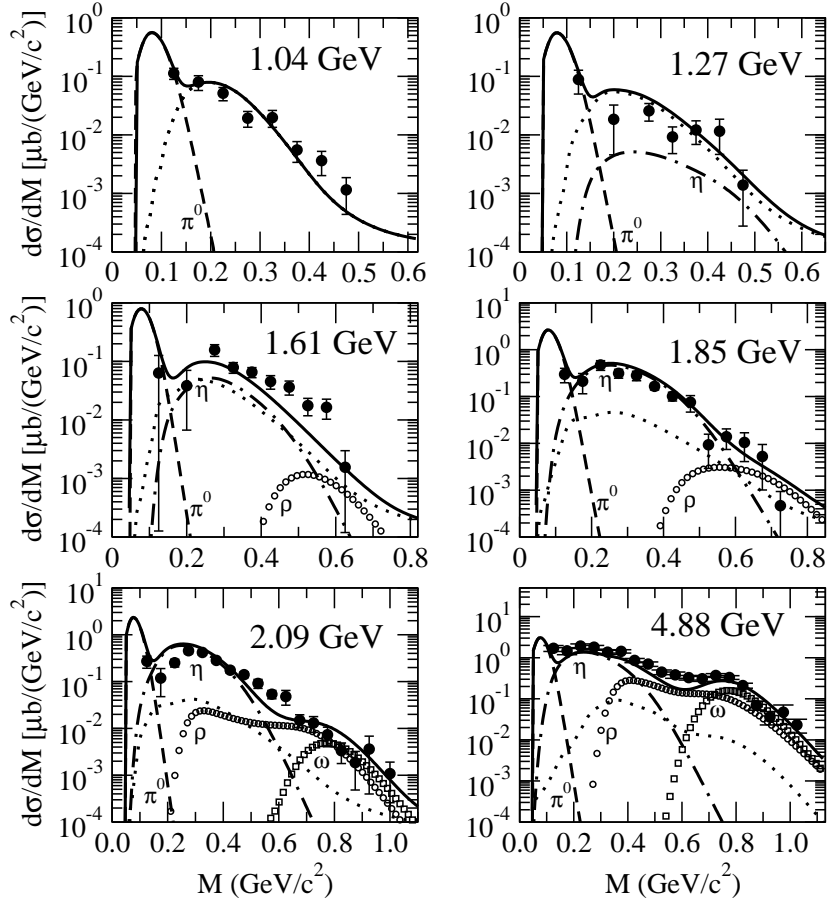}}
\end{center}
\vskip .1in
\caption {
The calculated dilepton invariant mass spectra for 
proton-proton collisions in the beam energy range of 1.04-4.88 GeV as
comparison to the DLS data~\protect\cite{wil98}. The effective
Lagrangian model results are shown by the dotted lines (the same
as those shown by full lines in Fig.~4. The contributions of $\pi^0$,
$\eta$ Dalitz decay and $\rho^0$ and $\omega$ direct decay processes
are shown by dashed, dashed-dotted, open circles, and
open squares, respectively. The incoherent sum of these
cross sections with those of the effective Lagrangian model is shown
by the solid line.  Solid circles represent the experimental data.
}
\label{fig:figf}
\end{figure}
\noindent
$M > $ 0.1 GeV. 
However, at beam energies higher than 1.27 GeV, our calculations fail
to reproduce the data, being lower than them by factors ranging from
2-20.  Clearly, with increasing beam energy, other dilepton sources
become important; these include Dalitz decays of hadrons
($\pi^0$, $\eta$ and $\omega$)~\cite{ern98,bra99,bra01}, 
direct decay of $\omega$ and $\rho^0$ meson to dileptons,
processes leading to multi-hadronic final states~\cite{hag94}
and two pion annihilation~\cite{kap89}.  Nevertheless, it is
encouraging that the present effective Lagrangian model is able to
account for the dilepton yields in $pp$ collisions at beam energies
around 1.0 GeV.  Therefore, this theory could be  useful in making
predictions for the dilepton spectra to be measured by the
HADES spectrometer at GSI, for beam energies in this region.
 
In order to study the role of other dilepton production
processes not considered in the present effective Lagrangian model,
we have included  the contributions of the Dalitz decays: 
$\pi^0 \rightarrow \gamma e^+e^-$ and $\eta \rightarrow \gamma e^+e^-$,
and the direct decays $\rho^0 \rightarrow e^+e^-$ and
$\omega \rightarrow e^+e^-$. The details of these calculations are 
described in Ref.~\cite{bra99} and important relevant formulas are listed in
appendix B.  The cross sections of these processes were incoherently
summed to those of the total mass differential cross sections (solid
lines in Fig.~5) of the ELM. The results are shown in Fig.~6 together
with the DLS data.  At the beam energies of 1.04 GeV and 1.27 GeV, we
have shown explicitly only the contributions of $\pi^0$ and $\pi^0$ and
$\eta$ decay processes, respectively together with the ELM results as cross
sections for the other decay channels are insignificant here. At 
incident energies of 1.61 GeV and 1.85 GeV we show in addition
the contributions of $\rho^0$ direct decay also; the $\omega$ direct decay 
is insignificant at these energies. However, at 2.09 GeV 
and 4.88 GeV, contributions of all the decay channels as mentioned
above are shown explicitely. 

It is clear that now the data can be described
reasonably well for all the beam energies. This comparison suggests
that the lowest points in the mass distributions of the dilepton stem
from the $\pi^0$ Dalitz decay. The $\eta$ Dalitz decay is important at the
intermediate masses for beam energies between 1.61 GeV to 4.88 GeV, while
$\rho^0$ and $\omega$ direct decay processes are important at higher
mass ends of the spectra for these beam energies. In addition to these
processes, the multi-hadron final state bremsstrahlung mechanism could
also contribute significantly in the low mass region at the beam energy
of 4.88 GeV~\cite{hag94}. We would like to mention that although
several processes are included in our results shown in Fig.~6, the
double counting problem is unlikely to be present there. 
The Dalitz decay and the direct decay processes of dilepton production
are excluded from the effective Lagrangian treatment. Furthermore, we 
have not considered the processes like $NN \rightarrow NR \rightarrow
NN\rho^0 \rightarrow NN e^+e^-$ (termed as the subthreshold resonance
production). Therefore, problems of double counting 
in the vector meson direct decay contributions as suggested in  
Ref.~\cite{fae03}, are not present in our calculations.

\section{Summary and conclusions}

We investigated the dilepton production in the nucleon-nucleon 
collisions at beam energies in the range of 1-5 GeV within an
effective Lagrangian model, which in proven
to describe well the
pion and kaon production in $NN$ collisions. Most of the parameters of
this model are fixed by fitting to the elastic $NN$ $T$ matrix; this
restricts the freedom of varying the parameters of the model to provide
a fit to the data. Along with the $NN$ bremsstrahlung process, the model
also includes the excitation, propagation and radiative decay of
$\Delta$(1232) and $N^*$(1520) intermediate nucleon resonant states.
The coupling constants at vertices involving resonances have been
determined from the experimental branching ratios of their decay into
various relevant channels. The interference 
terms among various amplitudes are included in the total $T$ matrix.

The reaction proceeds predominantly via excitation of the $\Delta$
intermediate state in the entire beam energy range considered in this
work. The contributions of $N^*$(1520) terms are relatively small.
The $NN$ bremsstrahlung contributions are also weak in comparison 
to those of the delta isobar except for the higher mass region of the 
invariant mass spectra  at the beam energy of 1.04 GeV. In this region
of these spectra considerable interference between various terms is
also visible. For the case of $pp$ collisions, the $NN$ bremsstrahlung and
$N^*$(1520) contributions are similar in magnitude at lower beam energies.
However, for the $pn$ case, the former is larger than the latter at
these energies. On the other hand, for the beam energies of
4.88 GeV, $N^*$(1520) terms are larger than those of the bremsstrahlung
in both the cases.  A key result of our study is that the $pn$
bremsstrahlung is stronger than the $pp$ one even at the beam energy
of 4.88 GeV. 

At the lower beam energies (1.04 GeV and 1.27 GeV), the effective
Lagrangian model calculations are able to explain the 
data of the DLS collaboration for the invariant mass distribution
of dilepton emitted in $pp$ collisions except for the
lowest mass point. This is encouraging in the context of the
analysis of the new experimental data on the dilepton production
expected shortly from the HADES collaboration at GSI, Darmstadt. 

However, at higher beam energies (1.61 GeV, 1.85 GeV, 2.09 GeV and 
4.88 GeV),  the effective Lagrangian model 
underpredicts the DLS data. This
indicates that with increasing energy, other dilepton production
sources become important. To stress this point further, we 
incoherently summed the $\pi^0$ and $\eta$ Dalitz decay and
$\rho^0$ and $\omega$ direct decay contributions to the  
cross sections of the ELM. With this procedure, it is 
possible to provide a good description of the DLS data
on the invariant mass distribution at all the beam energies. 
This also removes the discrepancy observed between the effective
Lagrangian model calculations and the DLS invariant mass distribution
data for the lowest mass point at the beam energies of 1.04 GeV and 1.27 GeV.    

Our study has clarified the role of $\Delta$ and $N^*$(1520)
intermediate baryonic resonance states in the dilepton production
in the $NN$ collisions. The interference effects of various terms are
visible towards the large invariant mass ends of the spectra for
proton-proton collisions. However, the DLS experimental filter
effectively cuts off this mass region. Therefore, these effects do not show 
up when calculations are compared with the data. This may imply that 
an incoherent combination of resonance and decay contributions is 
sufficient to explain the DLS data. However, with improved quality of 
the data this impression could change. Therefore, precise data  
where also the hadrons in the final channel are detected would be welcome.
This would put constraint on the models calculating  
the role of various mechanisms in the dilepton production
in nucleon-nucleon collisions. The present effective Lagrangian approach
should be improved by including the amplitudes for the decay 
processes also so that they can be summed coherently with the 
amplitudes of the other channels considered already.
\acknowledgements
This work has been supported by the Forschungszentrum J\"ulich.
One of the authors (RS) would like to thank
Elena Bratkovskaya for several useful and clarifying discussions
about the work presented in Refs.~\cite{bra99,bra01} and help in
implementing the filter program of the DLS collaboration. 

\appendix
\section{Some details of the phase space and the amplitude }

In this appendix we give additional details of the calculation of the
four-body phase space factor and the amplitude $A^{fi}$.

The kinematical situation for calculating the differential cross section 
[Eq.~(28)] is illustrated in Fig.~7 where ${\bf p_1}$ and ${\bf p_2}$ depict
the initial momenta of two nucleons while ${\bf p_3}$ and $ {\bf p_4}$ their
final momenta. The momenta of $e^+$ and $e^-$ are shown by 
${\bf p_+}$ and ${\bf p_-}$, respectively. 
{\bf P} represents half of the total center of mass (CM) dilepton momentum.
The differential cross section  can be written as
\begin{eqnarray}
\frac{d\sigma}{dM d\Omega_P dP d\Omega_{Q_L}d\Omega_{Q_N}} & = &
\frac{4m_N^4 m_e^2 P^2 Q_L^2 Q_N^2}{[E_{l+} Q_L-E_{l-} P cos \theta_1]
[(E_3 + E_4)Q_N+(E_3 - E_4)P cos \theta_2]}
 \nonumber \\ & \times &
\frac{M}{E_{l+} E_p|P_p|(2\pi)^8}\,\, |A^{fi}|^2,
\end{eqnarray}
where ${\bf Q}_L$ = ${\bf p}_-$ - ${\bf P}$ and ${\bf Q}_N$ = ${\bf p}_3$
+ ${\bf P}$.  $\theta_1$ and $\theta_2$ are angles 
between ${\bf P}$ and ${\bf Q}_L$, and ${\bf P}$ and ${\bf Q}_N$,
respectively. In Eq. (A1), we have defined $E_{l+} = E_+ + E_-$ 
and $E_{l-} = E_+ - E_-$, where $E_+$ and $E_-$ are energies of 
$e^+$ and $e^-$, respectively. $E_p$ and $|P_p|$ are the energy and 
momentum of the projectile nucleon in the CM system. The invariant
mass of the dilepton is given by, $M^2 = (p_+ + p_-)^2$.
Integrations over the residual degrees of freedom in the differential 
cross section [Eq.~(A1] are carried out by using the Monte Carlo techniques. 
\begin{figure}[here]
\begin{center}
\mbox{\epsfig{file=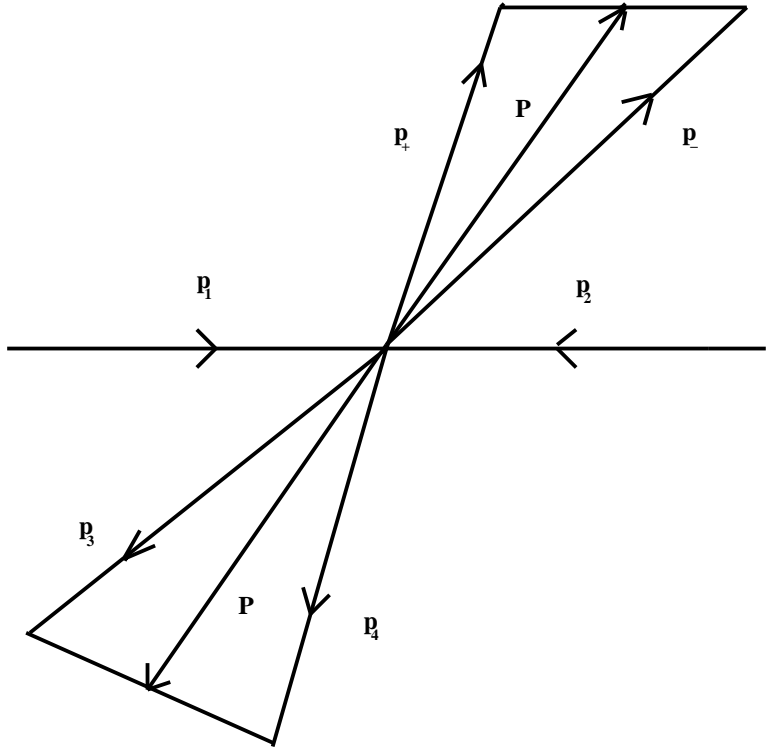,height=10cm}}
\end{center}
\vskip .1in
\caption {
Illustration of the kinematical situation for $e^+e^-$ pair production in
$NN$ collisions. ${\bf p_1}$ and ${\bf p_2}$ are the initial nucleon momenta
in the CM system while ${\bf p_3}$ and ${\bf p_4}$ are the same for the
final nucleons.  ${\bf P}$ is half of the total dilepton momentum. 
}
\label{fig:figg}
\end{figure}
\noindent

A total number of 90 diagrams [42 for $NN$ bremsstrahlung, 16 involving 
$\Delta$ intermediate states and 32 involving $N^*$(1520)] contribute to
the amplitude $A^{fi}$. Due to their large number, it is hard to calculate
them by usual trace techniques. Instead we follow the procedure,
given below. Similar method has also been used in our previous calculations
within this model.
 
We first note that amplitude corresponding to each individual diagram
can be split into a hadronic part and a leptonic part as 
\begin{eqnarray}
A^{fi} &\sim & h^{fi}_\mu \frac{g^{\mu\nu}}{k^2} l^{fi}_\nu,
\end{eqnarray}
where $k$ is the photon momentum.  For example,
for the diagram of type (a) of Fig.~1, involving pion exchange and the
nucleon intermediate state ($NN$ bremsstrahlung), we have 
\begin{eqnarray}
h_\mu^{fi} & = & -i\left( \frac{g_{NN\pi}}{2m_N} \right)^2 \frac{1}{q^2-m_\pi^2}
{\bar{u}}(p_4,s_4) \gamma_5 q\!\!\!/ u(p_2,s_2) {\bar{u}}(p_3,s_3)
\Gamma_\mu^{NN\gamma} \nonumber \\ & \times & 
\frac{(p_i\!\!\!\!\!/ + m_N)}{p_i^2-m_N^2}
\gamma_5 q\!\!\!/ u(p_1,s_1) f_{iso},\\
l_\nu^{fi} & = & e {\bar{u}}(p_+,s_+) {\hat{e}} \gamma_\nu u(p_-,s_-),
\end{eqnarray} 
where $u(p_i,s_i)$ are the free Dirac spinors with four momentum $p_i$
and spin $s_i$, and $\Gamma_\mu^{NN\gamma}$ is the same as defined in 
Eq.~(16). $f_{iso}$ is the isospin factor which for each graph, is obtained
from the separate treatment of the isospin parts. In Eq.~(A4)
${\hat{e}}$ is $\frac{1}{2}(1+\tau_3)$. 
Similar expressions for the amplitudes corresponding to all the graphs can 
be written in a straight forward way by using the Lagrangians and the
propagators given in the main text.

The summation over spins of the modulus square of the leptonic part 
gives
\begin{eqnarray}
L_{\mu\nu} & = & \sum_{spins} l_\mu^{fi\dagger}\l_\nu^{fi} \nonumber \\
& = & \frac{e^4}{4m_e^2}\left[ -2g_{\mu\nu}M^2 + 4 (p_{+\mu}
  p_{-\nu} + p_{+\nu} p_{-\mu}) \right ]
\end{eqnarray}
In the evaluation of the hadronic part of the  
amplitudes we encounter terms of the form
\begin{eqnarray}
T^{NI}_\mu & = & {\bar{u}}(p_3,s_3) \gamma_\mu \frac{p_{i\nu}\gamma^\nu + m_j}
{(p_i^2 - m_j^2)} \Gamma^{NI}_{eff} u(p_1,s_1),
\end{eqnarray}
where $\Gamma^{NI}_{eff}$ is a linear combination of 16 4 $\times$ 4 Dirac
matrices (1, $\gamma_\mu$, $\sigma_{\mu \nu}$, $\gamma_5$,
$\gamma_5 \gamma_\mu$)and $m_j$ is mass of either nucleon or the resonance.
In the latter case it also includes the resonance width as shown in 
Eq.~(24).
Further calculations of such terms are 
carried out by using the manipulator package REDUCE.  
The matrix elements so calculated are summed together as complex numbers.
The result still depends on the initial and final spins of the nucleons.
The modulus square of this sum is then averaged over the
initial spins and summed over the final spin states. This leads
to $|A^{fi}|^2$. By this procedure a large number of matrix manipulations
are avoided.
 
\section{Calculations of Dalitz decays and direct decay of mesons}

In this appendix we describe briefly the calculations  
of Dalitz decays $\eta \rightarrow \gamma e^+e^-$ and
$\pi^0 \rightarrow \gamma e^+e^-$, and direct decays 
$V \rightarrow e^+e^-$ ($V$ = $\rho$ and $\omega$).
More details can be found in Ref.~\cite{bra99}. 
 
The calculation of the process $pp \rightarrow M X \rightarrow \gamma e^+e^-$
proceeds in two steps. First the meson $M$ ($\eta$ and $\pi$) is produced in 
the $pp$ collisions and then the Dalitz decay $M \rightarrow \gamma e^+e^-$
is considered. In the procedure of Ref.~\cite{bra99}, the prametrizations
of Refs.~\cite{wol90,wol93,vet91} are used to calculate the $M$ production 
cross sections. The $\eta$ Dalitz decay to $\gamma e^+e^-$ is  
given by (see also~\cite{lan85})
\begin{eqnarray}
\frac{d\Gamma_{\eta\rightarrow \gamma e^+e^-}}{dM} & = &
\frac{4\alpha}{3\pi} \frac{\Gamma_{\eta\rightarrow 2\gamma}}{M}
\left( 1-\frac{4m_e^2}{M^2} \right)^{1/2}
\left( 1+2\frac{m_e^2}{M^2} \right)^{1/2} \nonumber \\ & \times &
\left( 1-\frac{M^2}{m_\eta^2} \right)^{1/2}
|F_{\eta\rightarrow \gamma e^+e^-} (M)|^2,
\end{eqnarray}
where the form factor is parameterized as
\begin{eqnarray}
F_{\eta\rightarrow \gamma e^+e^-} (M) & = &
\left( 1-\frac{M^2}{\Lambda_\eta^2} \right)^{-1}.
\end{eqnarray}
The value of parameter $\Lambda_\eta$ is 0.72 GeV.
In Eq.~(B1), $\alpha$ represents the
fine structure constant, and $\Gamma_{\eta \rightarrow 2\gamma}$
is the partial width for the $\eta \rightarrow 2\gamma$ decay
whose value is adopted from the particle data group 
compilation~\cite{PDG02}. 

The Dalitz decay $\pi^0 \rightarrow \gamma e^+e^-$
is  calculated in the same way using the following form factor 
\begin{eqnarray}
F_{\pi^0\rightarrow \gamma e^+e^-} (M) & = &
\left( 1 + B_{\pi^0} M^2 \right),
\end{eqnarray}
with $B_{\pi^0}$ = 5.5 GeV$^{-2}$. 

Similarly, calculations  of the direct decay of $\rho$ and $\omega$ mesons
proceed in two steps; the production of these meson in $pp$ collisions 
is calculated in the first step and their decay to dilepton in the
second 
\begin{eqnarray}
\frac{d\sigma (M)}{dM}^{NN \rightarrow VNN \rightarrow e^+e^-NN} 
& = & \frac{d\sigma (M)}{dM}^{NN \rightarrow VNN}
 \frac{\Gamma_{V \rightarrow e^+e^-}(M)}{\Gamma_{tot}^V (M)}
\end{eqnarray}
The calculation of the first term on the right hand side is described
in details in Ref.~\cite{bra99} and will not be repeated here. The 
decay width $\Gamma_{V \rightarrow e^+e^-} (M)$ is given by
\begin{eqnarray}
\Gamma_{V \rightarrow e^+e^-} (M) & = & C_V \frac{m^4_V}{M^3},
\end{eqnarray}
where $C_V$ = 8.8$\times$10$^{-6}$ and 0.767$\times$10$^{-6}$ for 
$V$ = $\rho$ and $\omega$, respectively. The total vector meson decay
width $\Gamma_{tot}^V$ is defined in the same way as in Ref.~\cite{bra99}.  
We refer to~\cite{fae03}, for a more recent description of these
calculations.  

\end{document}